\documentclass[%
 reprint,
 showkeys,
preprintnumbers,
 amsmath,amssymb,
 aps,
pra,
floatfix,
]{revtex4-2}

\usepackage{graphicx}
\usepackage{dcolumn}
\usepackage{bm}
\usepackage{physics}
\usepackage{color}
\usepackage{xcolor}

\begin{document}

\preprint{IFJPAN-IV-2025-24, P3H-25-093, TTK-25-39, ZU-TH~75/25}

\title{Higher-order QCD corrections to top-quark pair production in association with a jet}

\author{Simon Badger}
\email{simondavid.badger@unito.it}
\affiliation{Dipartimento di Fisica and Arnold-Regge Center, Università di Torino, and INFN, Sezione di Torino, Via P. Giuria 1, I-10125 Torino, Italy}
\author{Matteo Becchetti}
\email{matteo.becchetti@unibo.it}
\affiliation{Dipartimento di Fisica e Astronomia, Università di Bologna e INFN, Sezione di Bologna, via Irnerio 46, I-40126 Bologna, Italy}
\author{Colomba Brancaccio}
\email{colomba.brancaccio@unito.it}
\affiliation{Dipartimento di Fisica and Arnold-Regge Center, Università di Torino, and INFN, Sezione di Torino, Via P. Giuria 1, I-10125 Torino, Italy}
\author{Michal Czakon}
\email{mczakon@physik.rwth-aachen.de}
\affiliation{Institut f\"ur Theoretische Teilchenphysik und Kosmologie, RWTH Aachen University, D-52056 Aachen, Germany}
\author{Heribertus Bayu Hartanto}
\email{bayu.hartanto@apctp.org}
\affiliation{Asia Pacific Center for Theoretical Physics, Pohang, 37673, Korea, and Department of Physics, Pohang University of Science and Technology, Pohang, 37673, Korea}
\author{Rene Poncelet}
 \email{rene.poncelet@ifj.edu.pl}
\affiliation{Institute of Nuclear Physics PAN, ul.\ Radzikowskiego 152, 31-342 Krak\'ow}
\author{Simone Zoia}
\email{simone.zoia@physik.uzh.ch}
\affiliation{Physik-Institut, Universität Zürich, Winterthurerstrasse 190, 8057 Zürich, Switzerland}

\date{\today}

\begin{abstract}
The production of a top-quark pair, the heaviest known elementary particle, in association with a light jet is a key process for studying the properties of the Standard Model of Particle Physics. Due to its significance as a signal process with considerable sensitivity to the top-quark mass and as a background process for new physics searches, it is crucial to predict differential cross sections with high precision. In this article, we present, for the first time, predictions for various kinematical observables at next-to-next-to-leading order in Quantum Chromodynamics. The perturbative behavior is analyzed, and uncertainties arising from missing higher-order contributions are substantially reduced. The necessary two-loop amplitudes have been evaluated in the leading-color approximation, and we provide estimates for the impact of the missing contributions. 
\end{abstract}


\maketitle


\section{Introduction}
Hadroproduction of top-quark pairs plays a crucial role in the Standard Model (SM) precision program of the Large Hadron Collider (LHC).
First of all, the special role of the top quark in the SM as the heaviest particle makes it of paramount importance to know its mass as precisely as possible, due to the far-reaching consequences for the SM's consistency~\cite{deBlas:2022hdk} and phenomenology~\cite{Degrassi:2012ry}.
Secondly, precise knowledge of top-quark pair production cross sections is vital for SM measurements, such as the determination of parton distribution functions (PDFs)~\cite{Bailey:2019yze} and the strong coupling constant $\alpha_S$~\cite{CMS:2014rml, CMS:2019esx}.
And lastly, top-quark pair production through Quantum Chromodynamics (QCD) is an omnipresent background for searches for physics beyond the SM \cite{ATLAS:2021dqb}.
Due to the high center-of-mass energy available at the LHC, top-quark pairs, with a given invariant mass, are not only created at the production threshold but also in the presence of additional hard radiation with high rates to allow for measurements that are not statistically limited.
This kinematic regime can be viewed as a final state in which, in addition to the top-quark pair, a high-transverse-momentum jet is reconstructed.
Not only is this process an integral part of top-quark phenomenology, but it also presents a viable alternative to measure the top-quark mass~\cite{Alioli:2013mxa, Bevilacqua:2016jfk, Fuster:2017rev, Alioli:2022lqo}.
Corresponding experimental measurements of the top-quark pair plus jet process have been presented in Ref.~\cite{CMS:2014jya, ATLAS:2015pfy,ATLAS:2019guf, CMS-PAS-TOP-13-006}.

Theory predictions for this process have been obtained at next-to-leading order (NLO) in QCD~\cite{Dittmaier:2007wz, Dittmaier:2008uj} and electroweak theory (EW)~\cite{Gutschow:2018tuk}, in the stable-top-quark approximation at fixed order and matched to parton-showers \cite{Kardos:2011qa, Alioli:2011as, Hoeche:2014qda,  Hoche:2016elu}.
The NLO QCD corrections have also been incorporated for leptonic top-quark decays in the narrow-width approximation (NWA) \cite{Melnikov:2010iu, Melnikov:2011qx} and for the fully off-shell process \cite{Bevilacqua:2015qha, Bevilacqua:2016jfk, Bevilacqua:2017ipv}.
The remaining theory uncertainties for differential observables are a limiting factor in the experimental analyses, see for example Ref.~\cite{ATLAS:2019guf}.
In this work, we present for the first time next-to-next-to-leading order (NNLO) QCD corrections for the production of a stable top-quark pair in association with a jet.
The computation we present here is complete, except for subleading color and heavy-quark-loop contributions in the finite remainder of the two-loop matrix elements, which are unknown to date.
This computation has been made possible by recent advances in treating the rational coefficients and special functions in two-loop multi-scale computations \cite{Badger:2022hno, Badger:2024fgb, Badger:2024dxo}, which enabled the calculation of the leading-color partonic matrix elements in all required channels \cite{ttjamp:temp}.
The computation of these amplitudes was particularly challenging due to the presence of the top-quark mass, leading to elliptic functions in the Feynman integrals and to high polynomial degrees of the rational coefficients, which required advanced finite-field reconstruction techniques.

The contributions from real radiation are handled within the sector-improved residue-subtraction scheme using the \textsc{Stripper} C++ framework. This work is the first to use a NNLO QCD subtraction scheme for a pair of massive colored final-state state particles plus a colored parton in the final state.
It constitutes, therefore, another test to demonstrate the completeness of this subtraction scheme.

The article is organized as follows: after detailing the computational setup in section \ref{sec:setup}, we discuss the NNLO QCD corrections to a range of relevant LHC observables at 13 TeV in section \ref{sec:pheno}.
This includes a discussion of the implications of these corrections to future top-quark mass measurements.
We conclude and discuss future applications of the presented results in section \ref{sec:conclusion}.

\section{Computational setup}\label{sec:setup}

We are studying the cross-section for the process
\begin{align}
 pp \to t\bar{t}j + X \;,
\end{align}
where we are working with five massless quark flavors and one massive (top) quark ($m_t=172.5$ GeV), which is decoupled from the running of $\alpha_s$ \cite{Bernreuther:1981sg}.
The top quark is considered stable, and the mass is renormalized in the on-shell mass scheme.
We define the (differential) cross section in perturbation theory in collinear factorization through NNLO QCD:
\begin{align}
    \dd \sigma^{\rm LO} &= \alpha_s^3 \dd \hat{\sigma}^{(0)} \;, \\
    \dd \sigma^{\rm NLO} &= \alpha_s^3 \dd \hat{\sigma}^{(0)}+\alpha_s^4 \dd \hat{\sigma}^{(1)} \;, \\
    \dd \sigma^{\rm NNLO} &= \alpha_s^3 \dd \hat{\sigma}^{(0)}+\alpha_s^4 \dd \hat{\sigma}^{(1)}+\alpha_s^5 \dd \hat{\sigma}^{(2)}\;.
\end{align}
The cross sections are computed within the four-dimensional sector-improved residue subtraction scheme as formulated in Ref.~\cite{Czakon:2010td, Czakon:2014oma, Czakon:2019tmo}.
The implementation in C++ applies to arbitrary processes at the LHC.
It has been well tested through numerous calculations, including top-quark pair production and a range of 2-to-3 processes, with and without final-state masses~\cite{Chawdhry:2019bji, Chawdhry:2021hkp, Czakon:2021mjy, Hartanto:2022qhh, Badger:2023mgf}.

The computation requires the evaluation of tree-level amplitudes with up to 7 legs, which is handled within the AvH library~\cite{Bury:2015dla}.
The necessary six-point one-loop amplitudes are taken from  OpenLoops2~\cite{Cascioli:2011va, Buccioni:2019sur}, which provides a stable numerical evaluation~\cite{Buccioni:2017yxi}.
The two-loop finite-remainder for the partonic processes $gg\to t\bar{t}g$, $q\bar{q} \to t\bar{t}g$, and $qg \to t\bar{t}q$ are available in the strict leading color limit \cite{Badger:2022hno, Badger:2024fgb, Badger:2024dxo, ttjamp:temp}.
To minimize the impact of this approximation on the cross-section, we use the following reweighted definition of the finite remainder:
\begin{align}
F^{(2)}_{\rm l.c.\, resc.} (\mu^2) = \frac{F^{(0)}}{F^{(0)}_{\rm l.c.}} F^{(2)}_{\rm l.c.} (Q^2) + \sum_{i=0}^4 c_i \log^i(\mu^2/Q^2) \;,
\end{align}
where $Q^2$ is the partonic center of mass squared and the coefficients $c_i$ dependent on the lower-order finite remainders $F^{(0)}$ and $F^{(1)}$.
The finite remainders are defined as
\begin{align}
    F^{(0)} &= \langle \mathcal{F}^{(0)} | \mathcal{F}^{(0)} \rangle \;,\quad
    F^{(1)} = 2 \Re(\langle \mathcal{F}^{(0)} | \mathcal{F}^{(1)} \rangle) \;,\nonumber\\
    F^{(2)} &= 2 \Re(\langle \mathcal{F}^{(0)} | \mathcal{F}^{(2)} \rangle) + \langle \mathcal{F}^{(1)} | \mathcal{F}^{(1)} \rangle \;,\label{eq:finite}
\end{align}
through the $i$-th loop finite remainder amplitudes $| \mathcal{F}^{(i)} \rangle$.
The bra-ket notation indicates the spin and color-summed squares.
The infrared and ultraviolet renormalisation has been performed within the $\overline{\text{MS}}$ scheme, which is consistent with the IR scheme in Ref.~\cite{Czakon:2014oma}.

The numerical evaluation of the finite remainders in the leading color limit has been performed with the implementation in Ref.~\cite{ttjamp:temp}.
To efficiently evaluate the amplitude, we used an error-control strategy to solve the differential equations of the special functions, controlling the precision of the squared matrix element rather than that of the individual functions.
The contribution to the (differential) cross section has been evaluated by reweighting partially unweighted Born-level events.
This method has been validated at one-loop to reproduce the correct contributions to the differential cross section within statistical uncertainties.
The primary motivation for using this approach was better control over the numerical stability of the matrix elements.
The numerically stable evaluation of the matrix elements was a challenge due to cancellations in the rational functions of the amplitude.
All phase space points, in total about 90k across the studies presented here, have been evaluated in double-double (32-digit) floating point precision for the rational coefficients and double precision for the special functions.
However, the use of quad-double precision (64 digits) for the rationals was needed for about 4\% of the phase space points.

The required PDFs are evaluated using the LHAPDF library~\cite{Buckley:2014ana}, which is also used for the running of the strong coupling constant.
If not otherwise noted, we are using the PDF4LHC21 PDF set~\cite{PDF4LHCWorkingGroup:2022cjn}.
The cross sections are evaluated using a set of dynamical choices for the renormalization $\mu_R$ and factorization $\mu_F$ scales:
\begin{align} \label{eq:scales}
    \mu_R=\mu_F = \mu_0 = H_T/n \;,
\end{align}
with $n\in \{1,2,4\}$, where
\begin{align}
H_T = M_T(t)+M_T(\bar{t})+ p_T(j_1)\;.
\end{align}
Here, $M_T(t) = \sqrt{m_t^2+p_T(t)} $ is the transverse mass of the top-quark and $p_T(j_1)$ is the transverse momentum of the leading reconstructed $R=0.4$ anti-$k_T$ jet.
The remaining scale dependence is used to estimate the missing higher-order uncertainty (MHOU) via conventional seven-point scale variations by a factor of 2.
For comparison, we also estimate MHOU for the differential distributions via a theory-nuisance-parameter (TNP) approach \cite{Tackmann:2024kci, Lim:2024nsk, Cridge:2025wwo}.
In this method, contributions from the next perturbative orders are directly characterized in terms of a parametrization reflecting the structure of missing terms.
The parameterization comes with free parameters, and, together with a prior on their variation range, it allows for the estimation of the expected size of the missing contributions.
Specifically, we employ the parameterization described in Ref.~\cite{Lim:2024nsk}.

\begin{table}[t]
    \centering
    \begin{tabular}{l|c|c|c}
         & $\sigma^{\rm LO}$ [pb] & $\sigma^{\rm NLO}$ [pb] & $\sigma^{\rm NNLO}$ [pb] \\
        \hline

\multicolumn{4}{l}{$\mu_R = \mu_F = H_T/4$}\\
\hline
soft &  $341.9(0.2)^{+48.0\%}_{-30.2\%}$&  $383.0(1.3)^{+0.984\%}_{-9.9\%}$&  $362.9(10.1)^{+0.73\%}_{-4.27\%}$\\
hard &  $60.8(0.0)^{+49.0\%}_{-30.6\%}$&  $74.1(0.2)^{+4.27\%}_{-12.3\%}$&  $70.4(0.6)^{+0.593\%}_{-6.47\%}$\\
\hline
\multicolumn{4}{l}{$\mu_R = \mu_F = H_T/2$}\\
\hline
soft &  $238.5(0.1)^{+43.3\%}_{-28.3\%}$&  $345.1(0.8)^{+11.0\%}_{-14.4\%}$&  $363.0(6.2)^{+0.698\%}_{-4.73\%}$\\
hard &  $42.2(0.0)^{+44.1\%}_{-28.6\%}$&  $65.0(0.1)^{+14.1\%}_{-15.9\%}$&  $70.0(0.4)^{+1.08\%}_{-6.04\%}$\\
\hline
\multicolumn{4}{l}{$\mu_R = \mu_F = H_T$}\\
\hline
soft &  $171.1(0.1)^{+39.4\%}_{-26.6\%}$&  $295.5(0.6)^{+16.8\%}_{-16.2\%}$&  $345.8(4.1)^{+5.06\%}_{-7.97\%}$\\
hard &  $30.1(0.0)^{+40.1\%}_{-26.9\%}$&  $54.7(0.1)^{+18.9\%}_{-17.3\%}$&  $65.8(0.3)^{+6.43\%}_{-9.36\%}$\\
\hline
    \end{tabular}
    
    \vspace{.5cm}
    
    \includegraphics[width=0.45\textwidth]{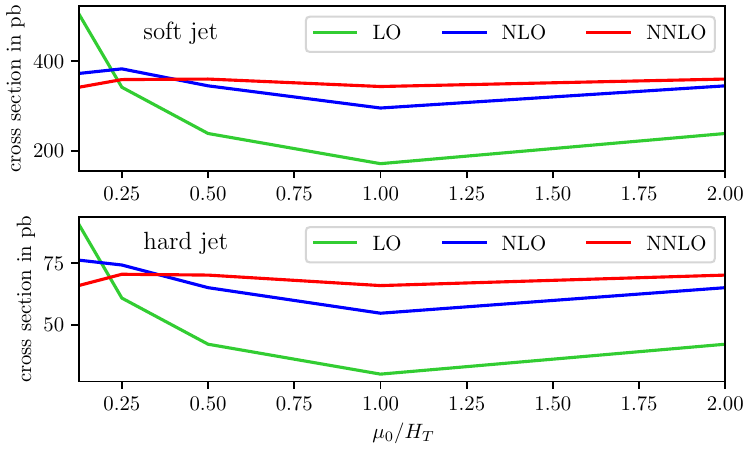}

    \caption{
    Fiducial cross sections in pb for $pp\to t\bar{t}j+X$ with different scale choices in the \textit{soft} and \textit{hard} phase space.
    The uncertainties are obtained from a seven-point scale variation around the central value and are expressed as relative percentages.
    The number in brackets represents the remaining statistical uncertainties on the central value.
    The plot shows the cross section as a function of $\mu_R = \mu_F =\mu_0$.
    }
    \label{tab:xsecs}
\end{table}

\section{LHC cross sections at NNLO QCD}\label{sec:pheno}
We study on-shell stable top-quark pair production at 13 TeV in proton-proton collisions.
While the inclusion of decays is planned for future work, this approximation is well justified given the high efficiency of reconstruction algorithms commonly used in experimental analyses.
We study two fiducial phase spaces for 13~TeV proton-proton collisions which are inclusive in the top-quark momenta:
\begin{enumerate}
    \item \textit{soft}: $p_T(j) \geq 30\; \text{GeV}$ and $|y(j)| \leq 2.5$,
    \item \textit{hard}: $p_T(j) \geq 120\;\text{GeV}$ and $|y(j)| \leq 2.4$, and additionally $p_T(j) \geq 0.4\cdot(p_T(t) + p_T(\bar{t}))$. These selection cuts are chosen according to ongoing studies in CMS \footnote{Private communication with Javier Llorente.}.
\end{enumerate}
The total cross sections for the different scale choices and phase spaces are shown in Tab.~\ref{tab:xsecs}.
The perturbative corrections behave essentially similarly in the \textit{soft} and \textit{hard} phase spaces.
The lowest scale choice ($H_T/4$) naturally yields the highest LO prediction, which is followed by mild NLO (about $11\%$) and slightly negative NNLO QCD ($-5\%$) corrections.
The highest scale choice ($H_T$) receives the most significant corrections ($+80\%$ at NLO and $+20\%$ at NNLO QCD).
The remaining scale dependence (by the overall range covered) at NLO (NNLO) QCD ranges from $10\%$ ($5\%$) with $\mu_0=H_T/4$ to $32\%$ ($12\%$) with $\mu_0=H_T$.
These observations indicate a faster perturbative convergence using smaller scales, similarly to the inclusive top-quark pair production case \cite{Czakon:2016dgf}.
Also similar to the inclusive top-quark pair production case is the asymmetry of the uncertainty using the $H_T/4$ choice.
For the differential distributions shown in the following subsection, we therefore settle for the $H_T/2$ choice.
The contribution from the double virtual corrections which are approximated using the leading-color two-loop finite remainder, i.e.\ $F^{(2)}_{\rm l.c.\, resc.} (\mu^2)$ in Eq.~\eqref{eq:finite} varies between 6--10\% of the total cross section depending on the scale choice.

\begin{figure}
    \centering
    \includegraphics[width=\linewidth]{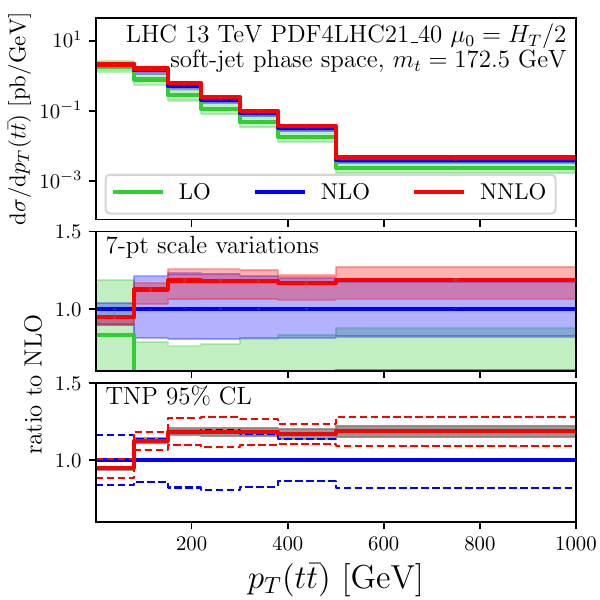}
    \caption{The transverse momentum distribution of the top-quark pair in the soft-jet phase space, through LO (green), NLO (blue), and NNLO (red) QCD. The top panel shows absolute predictions, while the bottom panels show ratios relative to the central NLO QCD prediction. The solid lines show the central predictions and the colored bands the envelope of seven-point scale variations. In the bottom panel, the dashed lines indicate the TNP uncertainty bands for MHOU, and the grey hashed area indicates the uncertainty estimates for the sub-leading color terms.}
    \label{fig:soft_pTtt}
\end{figure}

\subsection{Differential cross sections}

Next, we discuss differential cross sections. We will focus first on the transverse momentum distribution of the top-quark pair ($p_T(t\bar{t})$).

In figure~\ref{fig:soft_pTtt} we show the $p_T(t\bar{t})$ differential cross sections through LO, NLO, and NNLO QCD in the \textit{soft} phase space for the $H_T/2$ scale choice.
We observe the typical perturbative behavior, where higher-order corrections are successively smaller and the remaining sensitivity to the scales is reduced.
The NNLO QCD corrections are slightly ($-10\%$) negative in the low transverse momentum region and somewhat positive ($+10\%$) at high transverse momentum, indicating good perturbative convergence through the spectrum.
Uncertainties from TNPs (95\% CL) are similar in size to those from scale variations.
By construction, the TNP uncertainties are more symmetric than their estimates from scale variation.

In addition to the MHOU estimates, we use a simplified TNP-inspired approach to estimate the impact of missing sub-leading color terms on the differential spectrum.
To this end, we parameterize the sub-leading color terms in the following way:
\begin{align}
     \frac{F^{(2)}_{l.c.}}{F^{(0)}_{l.c.}} \to \frac{ F^{(2)}_{l.c.}}{F^{(0)}_{l.c.}} + \frac{\theta}{N_c^2}\frac{ F^{(2)}_{l.c.}}{F^{(0)}_{l.c.}} \;,
\end{align}
and estimate the impact on the differential cross sections by varying the parameter $\theta$ within $[-1,1]$.
The estimated uncertainty increases slightly with $p_T(t\bar{t})$ from about $1\%$ to $4\%$ in the high transverse momentum regime.

\begin{figure}
    \centering
    \includegraphics[width=\linewidth]{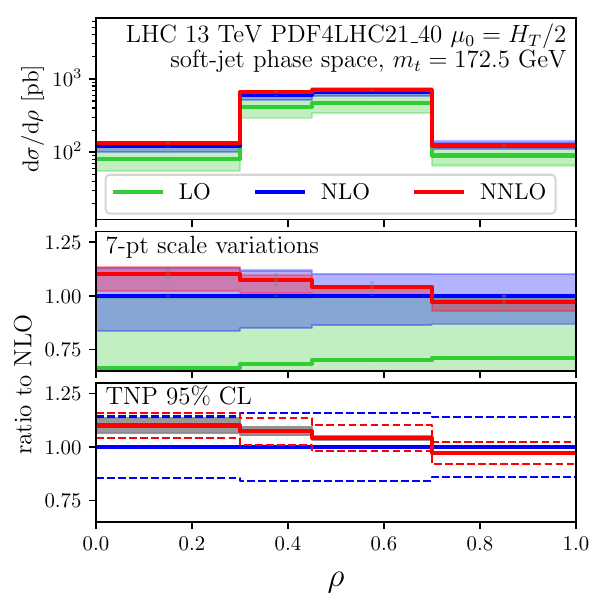}
    \caption{Same as Fig.~\ref{fig:soft_pTtt} but for the $\rho$ observable.}
    \label{fig:rho}
\end{figure}

The second observable we will discuss is
\begin{align}
\rho = \frac{2 m_0}{m(t\bar{t}j)} \;,
\end{align}
where $m_0=170\,\text{GeV}$ is an arbitrary scale \cite{Alioli:2013mxa}.
This observable provides an opportunity to extract the top-quark mass due to its sensitivity to the top-quark mass value; for a summary of NLO QCD studies, see Ref.~\cite{Alioli:2022lqo}.
In Fig.~\ref{fig:rho}, we show predictions through NNLO QCD and uncertainties from scale variation and TNPs as before.
The NNLO QCD corrections stabilize the perturbative expansion of the observable but also introduce a slight distortion in its shape, similar to those from NLO QCD corrections with respect to LO.
The uncertainties from scale variation and TNPs are consistent with each other.

\subsection{Cross section ratios}

\begin{figure}
    \centering
    \includegraphics[width=\linewidth]{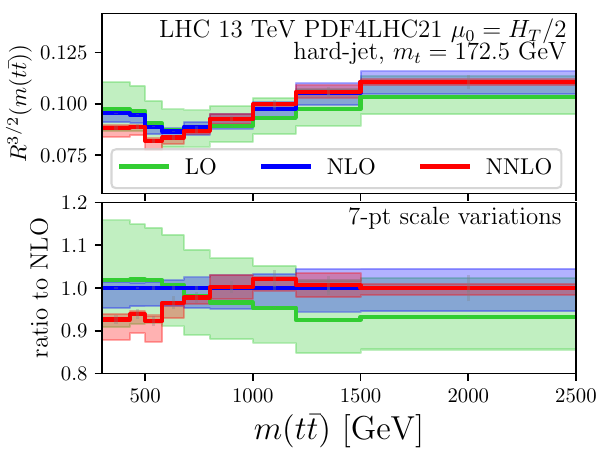}
    \caption{The R32 as a function of the top-quark pair invariant mass in the hard-jet phase space, through LO (green), NLO (blue), and NNLO (red) QCD. The top panel shows absolute predictions, while the bottom panel shows ratios relative to the central NLO QCD prediction. The solid lines show the central predictions and the colored bands the envelope of seven-point scale variations.
    }
    \label{fig:r32_mtt}
\end{figure}

We continue our discussion of differential observables by investigating the impact of higher-order corrections to cross-section ratios with respect to the inclusive top-quark pair production case, i.e., the cross section where no additional jet is required,
\begin{align}
    R^{3/2}(X) = \frac{\text{d} \sigma^{t\bar{t}j}(X)}{\text{d}\sigma^{t\bar{t}}(X)}\;,
\end{align}
as a function of an observable $X$ that is only based on the top-quark kinematics.
The factorization and renormalization scales for the top-quark pair cross sections are defined as in Eq.~\eqref{eq:scales} but without the jet transverse momentum.
We show the ratio as a function of the top-quark pair invariant mass in Fig.~\ref{fig:r32_mtt} in the \textit{hard} phase space for $\mu_0 = H_T/2$.
We refrain from providing uncertainties based on TNPs in this case, as both processes are clearly correlated and the TNP model in Ref.~\cite{Lim:2024nsk} does not provide an adequate correlation model.

At high invariant mass, the perturbative corrections to the ratio are small and reduce the MHOU by a factor of 2.
At low invariant mass, they come out larger and negative, reaching about $-7\%$ in the lowest bin.
This behavior indicates some perturbative instability.
Additionally, close to the $t\bar{t}$ threshold, Coulomb effects, which are of sub-leading color nature \cite{Barnreuther:2013qvf}, become important, which might limit the validity of the two-loop finite remainder leading color approximation.

\section{Conclusions}\label{sec:conclusion}

We presented predictions for cross sections for the production of top-quark pair in association with a reconstructed jet through NNLO QCD at the 13 TeV LHC.
The impact of higher-order corrections to differential observables relevant for LHC phenomenology suggests that NNLO QCD corrections are essential to achieve sub-10\% theoretical precision.
Besides the nominal impact, we studied estimates of theoretical uncertainties arising from missing higher orders due to scale variation and a TNP-based approach.
We found the expected substantial reduction in the remaining uncertainty (to about $\pm 10\%$), consistent across both methods, when including second-order corrections.
Also, we studied the impact of missing sub-leading color contributions arising from terms in the two-loop finite remainder, which add another $3-4\%$ uncertainty to the absolute NNLO QCD predictions at high transverse momentum or invariant mass.
Finally, we investigated ratios to inclusive top-quark pair production cross sections and found strong sensitivity to higher-order corrections and possibly sub-leading color effects at low invariant mass.

The present work represents only the starting point of a detailed investigation of the second-order QCD corrections to the top-quark pair plus jet process.
A crucial next step is to study the mass dependence in the $\rho$ observable at this order in perturbation theory.
Also the impact of the sub-leading color terms needs to be understood more rigorously.
A possibility might be offered in the high energy limit, where the top-quarks are quasi massless, where a comparison with (massified~\cite{Penin:2005eh, Mitov:2006xs, Becher:2007cu}) massless two-loop five-parton amplitudes~\cite{Agarwal:2023suw, DeLaurentis:2023izi, DeLaurentis:2023nss} could be performed, as these are known in full color.
A further future extension is the inclusion of top-quark decays within the NWA.
Finally, $t\bar{t}j$ NNLO QCD cross sections could be viewed as input for an N3LO QCD inclusive cross section computation within a slicing scheme, similar to recent achievements in color singlet production~\cite{Chen:2021isd, Chen:2021vtu, Chen:2022cgv, Chen:2022lwc}.

\begin{acknowledgments}
This work was performed in part using the Cambridge Service for Data Driven Discovery (CSD3), part of which is operated by the University of Cambridge Research Computing on behalf of the STFC DiRAC HPC Facility (www.dirac.ac.uk). The DiRAC component of CSD3 was supported by STFC grants ST/P002307/1, ST/R002452/1 and ST/R00689X/1.
The authors gratefully acknowledge the computing time provided
to them at the NHR Center NHR4CES at RWTH Aachen University (project number p0020025).
R.P.\ acknowledges that this research was funded in part by NCN 2024/55/D/ST2/00934.
S.B.\ and C.B.\ acknowledge funding from the Italian Ministry of Universities and Research (MUR) through FARE grant R207777C4R and through grant PRIN 2022BCXSW9.
M.B.~acknowledges funding from the European Union’s Horizon Europe research and innovation programme under the ERC Starting Grant No.~101040760 \emph{FFHiggsTop}.
M.C.\ acknowledges funding by the Deutsche Forschungsgemeinschaft (DFG) under grant 396021762-TRR 257: Particle Physics Phenomenology
after the Higgs Discovery.
H.B.H.~has been supported by an appointment to the JRG Program at the APCTP through the Science and Technology Promotion Fund and Lottery Fund of the Korean Government and by the Korean Local Governments~--~Gyeongsangbuk-do Province and Pohang City.
S.Z.~was supported by the Swiss National Science Foundation (SNSF) under the Ambizione grant No.~215960.
\end{acknowledgments}

\bibliography{ttbarj}

\end{document}